\documentclass[%
 reprint,
 amsmath,amssymb,
 aps,
 prl,
]{revtex4-2}

\usepackage{graphicx}
\usepackage{booktabs}
\usepackage{dcolumn}
\usepackage{bm}
\usepackage{xcolor}
\usepackage[colorlinks=true,allcolors=blue]{hyperref}

\newcommand{\bs}{\bm{s}}
\newcommand{\by}{\bm{y}}
\newcommand{\bxi}{\bm{\xi}}
\newcommand{\Tr}{\mathrm{Tr}}
\newcommand{\bu}{\bm{u}}
\newcommand{\hu}{\hat{\bm u}}
\newcommand{\C}{\mathsf C}
\newcommand{\R}{\mathsf R}
\newcommand{\U}{\mathsf U}
\newcommand{\Nn}{\mathsf N}
\newcommand{\RJ}{R_J}
\newcommand{\mt}[1]{Eq.~(\ref{#1}) of the main text}

\begin{document}

\title{Synchronous Monte Carlo method and its application to a mean-field spin glass model}

\author{Yoshiyuki Kabashima}
\altaffiliation[Also at ]{Trans-Scale Quantum Science Institute, The University of Tokyo, 7-3-1 Hongo, Bunkyo-ku, Tokyo 113-0033, Japan}
\email{kaba@phys.s.u-tokyo.ac.jp}
\affiliation{%
 Institute for Physics of Intelligence, The University of Tokyo,
 7-3-1 Hongo, Bunkyo-ku, Tokyo 113-0033, Japan
}%

\date{\today}

\begin{abstract}
Standard Markov chain Monte Carlo methods update variables one at a time to satisfy detailed balance, which prevents them from fully exploiting massively parallel hardware such as graphics processing units (GPUs). We study a Monte Carlo method for systems with pairwise interactions in which all variables are updated simultaneously, made possible by auxiliary Gaussian fields introduced through the Gaussian integral identity. The method satisfies detailed balance and is therefore guaranteed to converge to the canonical distribution. For mean-field spin glasses, its dynamics can be analyzed exactly by dynamical mean-field theory (DMFT). Applying the method and the DMFT to the random orthogonal model, which exhibits a random first-order transition, we find that the fluctuation-dissipation theorem (FDT) is clearly violated below the dynamical transition temperature, and that the relation between response and correlation takes the two-slope form of a generalized FDT, with simulations and theory in quantitative agreement.
\end{abstract}

\maketitle

\paragraph*{Introduction.}
Markov chain Monte Carlo (MCMC) methods are a basic tool of statistical physics~\cite{Metropolis1953,Glauber1963,Newman1999}, but they do not fit graphics processing units (GPUs) and similar accelerators well, which apply the same operation to a very large number of variables at once~\cite{Preis2009,Weigel2012}.
To satisfy detailed balance, the Metropolis and heat-bath (Glauber) rules update one variable at a time, conditioned on the current values of all the others.
For lattice models with short-range couplings, this restriction can be avoided by dividing the lattice into non-interacting sublattices (checkerboard decomposition)~\cite{Preis2009,Weigel2012,Janus2008}.
For models in which each variable interacts with many others, as in mean-field spin glasses, neural networks, and combinatorial optimization problems, no such decomposition exists, and the updates must remain sequential.
Updating all spins simultaneously with the heat-bath rule (Little dynamics~\cite{Little1974}) is easy to parallelize, but it violates detailed balance  
and its stationary distribution differs from the canonical one~\cite{Peretto1984}.

In this Letter we study an MCMC method for systems with pairwise interactions that updates all variables in parallel and still satisfies detailed balance~\cite{Martens2010,Zhang2012}.
The idea is to introduce auxiliary Gaussian fields through the Gaussian integral identity (the Hubbard--Stratonovich transformation~\cite{Stratonovich1957,Hubbard1959}).
Given the auxiliary fields, the spins are independent; given the spins, the auxiliary fields are Gaussian; both can therefore be sampled in parallel, with matrix--vector products and elementwise operations only.

Because the dynamics consists of a linear map followed by independent single-spin updates, it can be analyzed exactly for mean-field spin glasses with rotation-invariant couplings by dynamical mean-field theory (DMFT)~\cite{Sompolinsky1982,Eissfeller1992,Opper2016}.
We apply the method to the random orthogonal model (ROM)~\cite{Marinari1994,Parisi1995}.
Although this model is known to exhibit a random first-order transition (RFOT)~\cite{Kirkpatrick1987,Kirkpatrick1989,Cherrier2003,Caltagirone2012}, its dynamics---in particular, whether the fluctuation-dissipation theorem (FDT) is violated at low temperatures---has not, to our knowledge, been fully clarified.
Our simulations and DMFT analysis, which agree quantitatively, show that below the dynamical transition temperature this Ising model ages and clearly violates the FDT, and that the relation between response and correlation takes the two-slope form of a generalized FDT, as predicted for spherical $p$-spin models~\cite{Cugliandolo1993,Crisanti1993,Franz1998}.

\paragraph*{Algorithm.}
Consider $N$ Ising spins $\bs=(s_1,\dots,s_N)\in\{\pm1\}^N$ with energy
\begin{equation}
H(\bs)=-\frac{1}{2}\sum_{i,j}J_{ij}s_is_j-\sum_i h_is_i ,
\label{eq:H}
\end{equation}
where $J=(J_{ij})$ is a real symmetric matrix.
Since $s_i^2=1$, adding a constant $c$ to the diagonal of $J$ changes $H$ only by the constant $-cN/2$.
We choose $c\ge-\lambda_{\min}(J)$, where $\lambda_{\min}(J)$ is the smallest eigenvalue of $J$, so that $A\equiv\beta(J+cI)$ is positive semidefinite; here $\beta=1/T$ is the inverse temperature, with the Boltzmann constant set to unity.
The Gaussian integral identity
\begin{equation}
e^{\frac{1}{2}\bs^{\top}A\bs}
\propto\int d\bxi\,
\exp\!\left(-\frac{|\bxi|^2}{2}+\bxi^{\top}A^{1/2}\bs\right)
\label{eq:HS}
\end{equation}
shows that the canonical distribution $P(\bs)\propto e^{-\beta H(\bs)}$ is the marginal of the joint distribution
\begin{equation}
P(\bs,\bxi)\propto
\exp\!\left(-\frac{|\bxi|^2}{2}+\bxi^{\top}A^{1/2}\bs+\beta\bm{h}^{\top}\bs\right).
\label{eq:joint}
\end{equation}
Its two conditional distributions are simple.
Given $\bs$, $\bxi$ is Gaussian with mean $A^{1/2}\bs$ and unit covariance.
Given $\bxi$, the spins are independent, with $P(s_i\,|\,\bxi)=e^{s_iy_i}/(2\cosh y_i)$, where $\by=A^{1/2}\bxi+\beta\bm{h}$.
Alternating the two gives the following update, which we call synchronous Monte Carlo (SyncMC):
\begin{subequations}
\label{eq:SyncMC}
\begin{align}
\by^{t}&=A\bs^{t}+A^{1/2}\bm{z}^{t}+\beta\bm{h},
\label{eq:SyncMC_y}\\
s_i^{t+1}&=\pm1\ \text{with probability}\ \frac{1\pm\tanh y_i^{t}}{2}\quad(\text{all }i),
\label{eq:SyncMC_s}
\end{align}
\end{subequations}
where $\bm{z}^t$ is a vector of independent standard normal variables drawn afresh at each step.
Compared with Little dynamics, $s_i^{t+1}\sim e^{\beta s_i(\sum_jJ_{ij}s_j^t+h_i)}$, Eq.~\eqref{eq:SyncMC} contains two corrections: the diagonal shift $c$ and the correlated Gaussian noise $A^{1/2}\bm{z}^t$.
Together they remove the bias of the synchronous update.
Each step consists only of matrix--vector products and elementwise operations.
With enough processors it therefore takes a parallel time (span)~\cite{Blelloch1996} of order $\log N$, whereas a sweep of sequential updates takes a time of order $N$; the advantage, a factor $N/\log N$, grows with the system size [Fig.~\ref{fig:fig1}(b); Sec.~S1 of the Supplemental Material (SM)~\cite{SM}].
\begin{figure}[t]
\centering
\includegraphics[width=\linewidth]{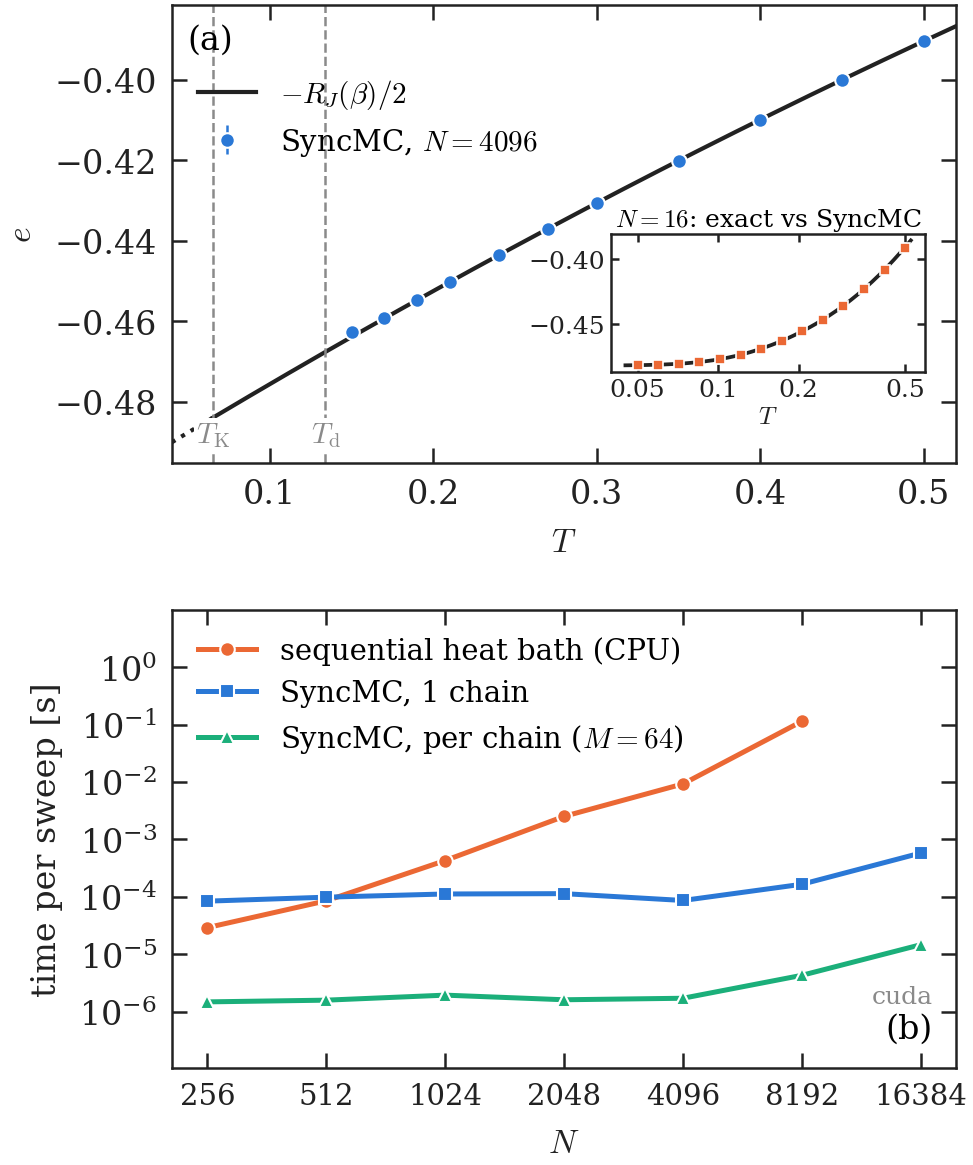}
\caption{(a) Energy per spin $e$ versus temperature $T$.
Circles: SyncMC for $N=4096$, annealed from high $T$ (error bars are smaller than the symbols).
Solid line: $e(T)=-R_J(\beta)/2$, the equilibrium energy for $T>T_{\rm K}$; dotted line: its continuation below $T_{\rm K}$.
Vertical dashed lines mark $T_{\rm K}$ and $T_{\rm d}$.
Inset: $N=16$; SyncMC with replica exchange (squares) versus exact enumeration (line).
(b) Wall-clock time per sweep ($N$ spin updates) versus $N$: sequential heat-bath updates on one CPU thread, and SyncMC on a GPU for a single chain and per chain when $M=64$ chains run together (conditions in SM Sec.~S1).}
\label{fig:fig1}
\end{figure}

\emph{Detailed balance.}---%
The transition probability of the spins in one step of Eq.~\eqref{eq:SyncMC} is $W(\bs'|\bs)=\int d\bxi\,P(\bxi|\bs)P(\bs'|\bxi)$.
Multiplying by $P(\bs)$ gives
\begin{equation}
P(\bs)W(\bs'|\bs)=\int d\bxi\,P(\bxi)\,P(\bs|\bxi)\,P(\bs'|\bxi),
\label{eq:DB}
\end{equation}
which is symmetric in $\bs$ and $\bs'$~\cite{Liu1994}.
Thus $W$ satisfies detailed balance with respect to the canonical distribution.
Since $W(\bs'|\bs)>0$, the chain converges to the canonical distribution from any initial condition, for any symmetric $J$, $N$, and $c\ge-\lambda_{\min}(J)$.

\paragraph*{Model.}
The ROM~\cite{Marinari1994,Parisi1995} is defined by Eq.~\eqref{eq:H} with $J=O\Lambda O^{\top}$, where $O$ is an $N\times N$ random orthogonal matrix drawn uniformly (Haar measure) and $\Lambda$ is diagonal with $N/2$ entries $+1$ and $N/2$ entries $-1$.
Its equilibrium behavior is described by the one-step replica symmetry breaking (1RSB) scenario: ergodicity is broken at the dynamical transition $T_{\rm d}=0.1336$ (plateau height $q_{\rm d}=0.961$), and the static transition occurs at $T_{\rm K}\simeq0.065$~\cite{Marinari1994,Cherrier2003}.
For $T>T_{\rm K}$, where the equilibrium free energy coincides with that of the paramagnetic phase, the energy per spin is $e(T)=-\frac{1}{2}R_J(\beta)$~\cite{Parisi1995}, where
\begin{equation}
R_J(z)=\frac{\sqrt{1+4z^2}-1}{2z}
\label{eq:RJ}
\end{equation}
is the R-transform of the eigenvalue distribution of $J$.
The same function determines the DMFT below.

Because the distribution of $J$ is invariant under rotations, the average of the dynamics over $O$ can be carried out for $N\to\infty$~\cite{Opper2016,Cakmak2019}, and the dynamics is then characterized by two quantities: the two-time correlation $C(t,t')\equiv N^{-1}\sum_i\langle s_i^ts_i^{t'}\rangle$ and the response $R(t,t')\equiv N^{-1}\sum_i\partial\langle s_i^t\rangle/\partial h_i^{t'}$ to a small field $h_i^{t'}$ applied only in the update $t'\to t'+1$.
Here $\langle\cdots\rangle$ denotes the average over the trajectories for a given $J$ (over the initial configuration, $\bm z^t$, and the spin updates); for $N\to\infty$ the results do not depend on the realization of $O$.
Because the field enters Eq.~\eqref{eq:SyncMC_s} only through $y_i^{t'}$, the response can be measured in the simulations exactly without applying any field~\cite{Chatelain2003,RicciTersenghi2003} (SM Sec.~S2):
\begin{equation}
R(t,t')=\frac{\beta}{N}\sum_i\left\langle s_i^t\left(s_i^{t'+1}-\tanh y_i^{t'}\right)\right\rangle .
\label{eq:response}
\end{equation}
We also use the integrated response $\chi(t,t_w)\equiv\sum_{t'=t_w}^{t-1}R(t,t')$.

\paragraph*{Dynamical mean-field theory.}
We average over $O$ with the generating-functional method~\cite{Martin1973,DeDominicis1978} combined with the asymptotics of spherical integrals~\cite{GuionnetMaida2005,Opper2016} (SM Sec.~S4).
In terms of the local field $u^t\equiv y^t/\beta$, each spin then follows the effective single-site process
\begin{subequations}
\label{eq:DMFT}
\begin{align}
u^t&=c\,s^t+\sum_{\tau<t}\big[R_J(\mathsf R)\big]_{t\tau}s^\tau+\eta^t,\\
s^{t+1}&=\pm1\ \text{with probability}\ \frac{1\pm\tanh\beta u^t}{2},
\end{align}
\end{subequations}
where $\eta^t$ is a Gaussian process with zero mean and covariance
\begin{align}
\mathsf N={}&\sum_{n\ge1}a_n\sum_{k=0}^{n-1}\mathsf R^k\,\mathsf C\,(\mathsf R^{\top})^{n-1-k}\nonumber\\
&+\frac1\beta\sum_{n\ge1}a_n\sum_{k=0}^{n}\mathsf R^k(\mathsf R^{\top})^{n-k}+\frac c\beta I ,
\label{eq:Ncov}
\end{align}
with $R_J(z)=\sum_{n\ge1}a_nz^n$ [for the ROM, $R_J(z)=z-z^3+2z^5-\cdots$].
The average over the rotation thus replaces the coupling term by a memory term acting on the past spins of the same site (a retarded Onsager reaction) and a Gaussian noise.
Both are fixed by the correlation matrix $\mathsf C_{tt'}=C(t,t')=\langle s^ts^{t'}\rangle$ and the response matrix $\mathsf R_{t\tau}=R(t,\tau)$, obtained in the single-site process with the estimator of Eq.~\eqref{eq:response}.
Since $\mathsf R$ is strictly lower triangular, the series in Eq.~\eqref{eq:Ncov} are finite (SM Sec.~S4), and every quantity at time $t$ depends only on earlier times, so the DMFT is solved in a single pass forward in time, with no iteration: at each step, row $t$ of $\mathsf C$ and $\mathsf R$ is estimated from many sampled trajectories of the single-site process, as in Refs.~\cite{Eissfeller1992,Eissfeller1994}, and $\eta^t$ and $s^{t+1}$ are sampled.
Statistical errors of the kernels feed back into later times; we therefore average independent runs and use their spread as the error of the DMFT curves, which we computed with an equivalent representation of Eq.~\eqref{eq:DMFT} for the ROM (SM Sec.~S4).
The solution describes the $N\to\infty$ limit of the simulation for any initial condition, including the quench studied below, and the FDT can be checked directly on it.

\emph{Stationary and aging solutions.}---%
In equilibrium, detailed balance reduces Eq.~\eqref{eq:response} to the discrete-time FDT (SM Sec.~S2),
\begin{equation}
T\,R(t,t')=C(t,t'+1)-C(t,t'),
\label{eq:FDT_discrete}
\end{equation}
for any $N$; summing over $t'$ gives
\begin{equation}
T\chi(t,t_w)=1-C(t,t_w).
\label{eq:FDT}
\end{equation}
Equations~\eqref{eq:DMFT}, on the other hand, describe transients as well and do not imply the FDT by themselves; it holds for the stationary solution, in which $C$ and $R$ depend only on $t-t'$ (SM Sec.~S5).
Above $T_{\rm d}$ this solution has $C\to0$ as $t-t'\to\infty$, and the solution after a quench approaches it at long times.
Below $T_{\rm d}$ the dynamics started from a random configuration never becomes stationary.
For $t'\to\infty$ we assume a hierarchy of three time scales:
(i) $t-t'=O(1)$, where $C(t,t')$ decays from $1$ toward $q_{\rm EA}$ and depends only on $t-t'$;
(ii) $1\ll t-t'\ll t'$, where $C(t,t')$ stays at the plateau $q_{\rm EA}$;
and (iii) $t-t'=O(t')$, where $C(t,t')$ decays from $q_{\rm EA}$ to zero and depends on $t$ and $t'$ only through $\ln h(t)-\ln h(t')$ for some effective age $h(t)$ [$h(t)=t$ for simple aging], i.e., the dynamics is stationary with respect to $\lambda=\ln h(t)$.
On scale (i) the FDT holds; on scale (iii), self-consistency requires
\begin{equation}
T\,R(t,t')=x\,[C(t,t'+1)-C(t,t')]
\label{eq:GFDT_diff}
\end{equation}
with a constant $0<x<1$, instead of Eq.~\eqref{eq:FDT_discrete}~\cite{Cugliandolo1993,Crisanti1993}.
Summing Eqs.~\eqref{eq:FDT_discrete} and \eqref{eq:GFDT_diff} over $t'$ gives the generalized FDT for the integrated response,
\begin{equation}
T\chi(t,t_w)=
\begin{cases}
1-C(t,t_w), & C>q_{\rm EA},\\
1-q_{\rm EA}+x\,(q_{\rm EA}-C(t,t_w)), & C<q_{\rm EA},
\end{cases}
\label{eq:GFDT}
\end{equation}
so that the slow degrees of freedom respond as if at an effective temperature $T/x>T$~\cite{Cugliandolo1993,Crisanti1993,Cugliandolo1997}.
Under these assumptions, the slow part of the local field has the same distribution as the frozen field of the 1RSB solution, with the Parisi parameter replaced by $x$; $x$ is then fixed by requiring the plateau to be marginally stable against the fast dynamics, which gives $q_{\rm EA}=0.981$ and $x=0.679$ at $T=0.1$ (SM Sec.~S6).
The numerical solution of Eqs.~\eqref{eq:DMFT} makes none of these assumptions, so comparing it and the simulations with Eqs.~\eqref{eq:FDT_discrete} and \eqref{eq:GFDT} tests them directly.

\paragraph*{Results.}
In all simulations we use the minimal shift $c=1$ unless stated otherwise (the SyncMC update for the ROM is given in SM Sec.~S3).
Figure~\ref{fig:fig1}(a) confirms that SyncMC samples the canonical distribution: for $N=16$, with replica exchange, it reproduces the exact energy down to $T=0.05<T_{\rm K}$, and for $N=4096$ the energy agrees with $e(T)=-R_J(\beta)/2$ above $T_{\rm d}$.
The time per SyncMC step on a GPU hardly grows with $N$ [Fig.~\ref{fig:fig1}(b)]; for $N=8192$ it is $10^2$ times shorter than a sequential sweep on a CPU ($10^4$ per chain for $64$ chains; conditions in SM Sec.~S1).

\begin{figure}[t]
\centering
\includegraphics[width=\linewidth]{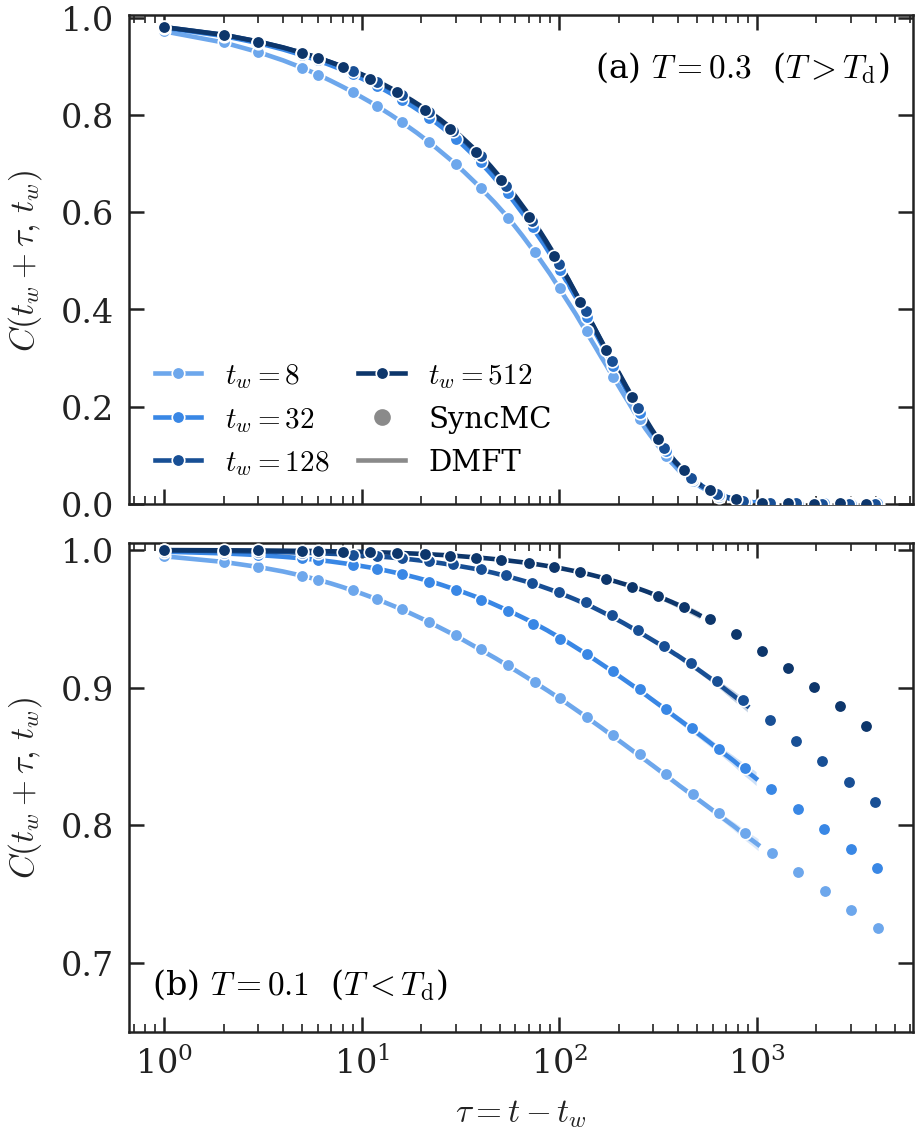}
\caption{Two-time correlation $C(t_w+\tau,t_w)$ after a quench from a random configuration, for $t_w=8$, $32$, $128$, and $512$ (light to dark).
Symbols: SyncMC for $N=16384$, averaged over $4$ samples of $O$ with $16$ independent chains each ($64$ chains in total); the standard errors are smaller than the symbols.
Lines: DMFT, averaged over $4$ independent runs with $2^{19}$ sampled trajectories each; the shaded band (mostly hidden by the lines) shows the spread over the runs.
(a) $T=0.3>T_{\rm d}$; (b) $T=0.1<T_{\rm d}$.}
\label{fig:fig2}
\end{figure}

Figure~\ref{fig:fig2} shows the two-time correlation $C(t_w+\tau,t_w)$ after a quench from a random configuration to temperature $T$, for several waiting times $t_w$.
Above $T_{\rm d}$ [Fig.~\ref{fig:fig2}(a)], the curves for $t_w\ge32$ coincide, i.e., $C$ depends only on $\tau$, and $C$ decays to zero: the system reaches equilibrium.
Below $T_{\rm d}$ [Fig.~\ref{fig:fig2}(b)], $C$ first stays close to $1$ and then decays on a time scale that grows with $t_w$: the older the system, the slower it relaxes, the hallmark of aging.
In both cases the DMFT curves pass through the SyncMC data over the whole window computed ($\tau\lesssim10^3$), whose standard errors are smaller than the symbols; finite-size effects are therefore negligible on these time scales.

\begin{figure}[t]
\centering
\includegraphics[width=\linewidth]{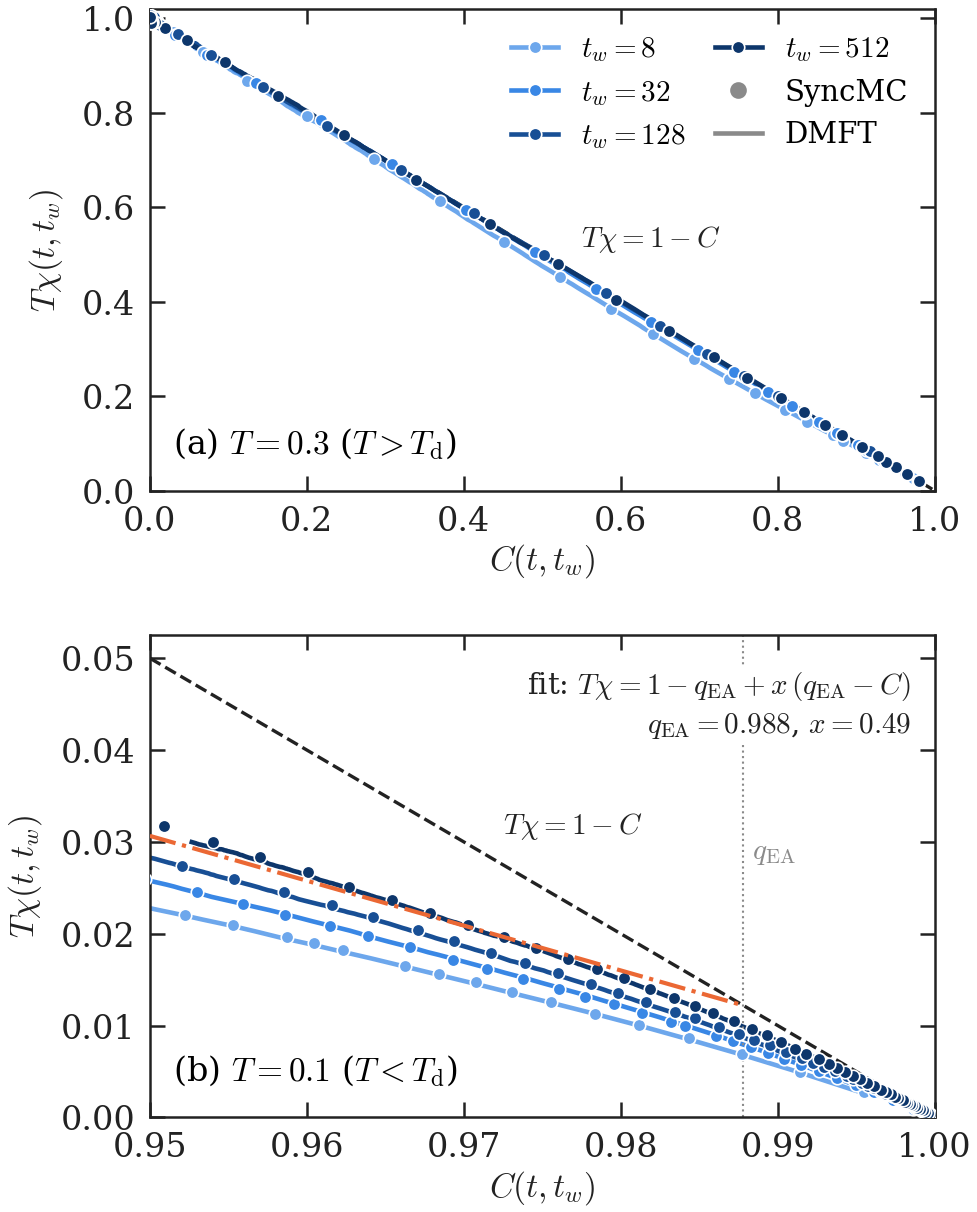}
\caption{Fluctuation-dissipation plot: $T\chi(t,t_w)$ versus $C(t,t_w)$ at fixed $t_w$; symbols (SyncMC), lines (DMFT), and colors as in Fig.~\ref{fig:fig2}.
Dashed line: the FDT, Eq.~\eqref{eq:FDT}.
(a) $T=0.3$.
(b) $T=0.1$, shown for $C\ge0.95$; the dash-dotted line is a fit of the generalized FDT~\eqref{eq:GFDT} to the SyncMC data for $t_w=512$, which gives $q_{\rm EA}=0.988$ and $x=0.49$.
The marginally stable 1RSB solution gives $q_{\rm EA}=0.981$ and $x=0.679$.}
\label{fig:fig3}
\end{figure}

The central result is the parametric plot of $T\chi(t,t_w)$ against $C(t,t_w)$ at fixed $t_w$ (Fig.~\ref{fig:fig3}).
Above $T_{\rm d}$ [Fig.~\ref{fig:fig3}(a)], the data fall on the line $T\chi=1-C$: the FDT~\eqref{eq:FDT} holds once the system has equilibrated.
Below $T_{\rm d}$ [Fig.~\ref{fig:fig3}(b)], the data follow this line only for $C$ very close to $1$, the fast relaxation within a metastable state; once $C$ falls below about $0.99$, they leave it and continue along a second, nearly straight line of smaller slope: the FDT is clearly violated, and the data take the two-slope form of Eq.~\eqref{eq:GFDT}.
SyncMC and DMFT agree throughout; since neither assumes stationarity in $\ln t$, this supports the aging solution.
A fit of Eq.~\eqref{eq:GFDT} to the data for $t_w=512$ gives $q_{\rm EA}=0.988$ and $x=0.49$, still below the marginal values expected for asymptotic aging~\cite{Cugliandolo1993,Castellani2005}.
Longer SyncMC runs ($N=4000$, up to $t=3.2\times10^6$) give slopes consistent with it, e.g., $x=0.68\pm0.10$ at $T=0.1$ and $0.79\pm0.09$ at $T=0.12$ (predicted $0.850$) for $t_w=10^5$, although the errors do not resolve a trend with $t_w$ (SM Sec.~S7).

\paragraph*{Discussion.}
Although SyncMC differs from the sequential dynamics usually considered in theory, the picture found here agrees with the 1RSB predictions, suggesting that it reflects the free-energy landscape rather than a particular dynamics; whether the asymptotic $q_{\rm EA}$ and $x$ are also independent of the dynamics remains open.
The diagonal shift $c$ does not change the stationary distribution but slows the relaxation, since it acts as a self-field of strength about $\beta c$ favoring the current sign: at $T=0.3$ the relaxation time grows by $11\%$ and $46\%$ for $c=1.05$ and $1.2$ (SM Sec.~S7); the minimal shift is therefore the best choice, and for it the gain in span, $N/\log N$, wins for large systems.
The method applies to any symmetric $J$ and combines readily with replica exchange~\cite{Hukushima1996}, and the DMFT requires only the R-transform of $J$, so the analysis extends directly to other rotation-invariant ensembles such as the SK model.

In summary, SyncMC updates all spins in parallel while satisfying detailed balance, and for rotation-invariant mean-field spin glasses its dynamics is described exactly by a DMFT.
Applied to the ROM, simulation and theory agree quantitatively and show that below the dynamical transition this Ising model ages and violates the FDT, with a response-correlation relation of the generalized-FDT form.
This differs from earlier studies of aging in Ising spin glasses~\cite{Yoshino1997,Parisi1999,Montanari2003} in that the theory is exact for $N\to\infty$ and describes the same detailed-balance dynamics that is simulated; the two-slope form, established analytically for spherical models~\cite{Cugliandolo1993,Crisanti1993}, is thus obtained for discrete spins and a Markov chain that samples the canonical distribution.

\begin{acknowledgments}
Generative-AI tools (GPT-5.5 and GPT-5.6 Sol, OpenAI; Claude Opus 5.5 and Claude Fable 5.1, Anthropic) were used extensively, under the author's direction, for calculations, code, simulations, data analysis, and drafting the text. The author checked the derivations and results, including by comparing theory with simulation, and takes full responsibility for the content of the manuscript.
This work was supported by JSPS KAKENHI Grant Nos.~22H05117 and 26K02981.
\end{acknowledgments}

\bibliography{SyncMC_ROM_arxiv}

\clearpage
\onecolumngrid
\begin{center}
{\large\bfseries Supplemental Material}
\end{center}
\setcounter{equation}{0}\setcounter{figure}{0}\setcounter{table}{0}\setcounter{section}{0}
\renewcommand{\theequation}{S\arabic{equation}}
\renewcommand{\thefigure}{S\arabic{figure}}
\renewcommand{\thetable}{S\arabic{table}}
\renewcommand{\thesection}{S\arabic{section}}
\renewcommand{\thesubsection}{\thesection.\Alph{subsection}}
\setcounter{secnumdepth}{2}
\makeatletter\renewcommand{\p@subsection}{}\makeatother

\noindent
Equation, figure, and reference numbers without the prefix S refer to the main text.
Sections~\ref{sec:cost}--\ref{sec:rom} collect technical details of the algorithm.
Section~\ref{sec:dmft} derives the dynamical mean-field theory (DMFT) for a general rotation-invariant coupling matrix from the generating functional of the dynamics.
Sections~\ref{sec:tti} and \ref{sec:aging} analyze its long-time solutions above and below $T_{\rm d}$, assuming the hierarchy of time scales stated in the main text.
Section~\ref{sec:num} contains additional numerical results.

\section{Computational cost}
\label{sec:cost}
One step of \mt{eq:SyncMC} consists of two matrix--vector products and $N$ independent single-spin updates.
Since \mt{eq:HS} holds with $A^{1/2}$ replaced by any $F$ satisfying $FF^{\top}=A$, a known eigenvalue decomposition $J=O\Lambda O^{\top}$ allows the choice $F=OD^{1/2}$ with $D=\beta(\Lambda+cI)$:
\begin{equation}
\bxi^t=D^{1/2}O^{\top}\bs^t+\bm{z}^t,\qquad \by^t=OD^{1/2}\bxi^t+\beta\bm{h}.
\label{eq:SyncMC_eig}
\end{equation}
The auxiliary field is then generated component by component, in $O(N)$ operations, and no matrix square root is needed; if only $K$ eigenvalues of $D$ are nonzero, $O$ is replaced by its $N\times K$ block.
Every operation is thus a matrix--vector product or an elementwise operation.
With enough processors, one step therefore takes a parallel time (span)~\cite{Blelloch1996}, i.e., the length of the longest chain of operations that must be done one after another, of order $\log N$, the time needed to sum $N$ numbers, whereas a sweep of sequential updates takes a time of order $N$ however many processors are used, because the spins must be updated one after another.
The advantage, a factor $N/\log N$, grows with the system size.
Running $M$ replicas together turns the products into matrix--matrix products, which GPUs handle most efficiently.

The wall-clock times in Fig.~\ref{fig:fig1}(b) of the main text were measured as follows.
SyncMC ran on one GPU (NVIDIA H200) in single precision (float32), with $c=1$ and $\beta=5$, for a single chain and for $M=64$ chains run together (the latter converted to time per chain); after 5 warm-up steps, the time was averaged over 50 steps, for $N=256$--$16384$.
The sequential heat-bath updates ran on one thread of a CPU (Intel Xeon 8468) in double precision (float64), compiled with numba, at $\beta=5$ starting from a random configuration; after compilation, the time was averaged over 5 sweeps, for $N=256$--$8192$.
Since the two methods ran on different hardware and in different precision, the ratio of their times is an indication only; the point of the figure is that the time per SyncMC step hardly grows with $N$, as expected from its span.

\section{Response estimator and the FDT}
\label{sec:response}
Let $h_i^{t'}$ be a field applied only in the update $t'\to t'+1$, so that $y_i^{t'}$ in \mt{eq:SyncMC_s} is replaced by $y_i^{t'}+\beta h_i^{t'}$.
The probability of the path depends on $h_i^{t'}$ only through the factor $e^{s_i^{t'+1}(y_i^{t'}+\beta h_i^{t'})}/[2\cosh(y_i^{t'}+\beta h_i^{t'})]$, whose logarithmic derivative at $h=0$ is $\beta(s_i^{t'+1}-\tanh y_i^{t'})$.
Differentiating $\langle s_i^t\rangle$ for $t>t'$ therefore gives \mt{eq:response}.

In equilibrium, the pair $(\bs^{t'},\by^{t'})$ is distributed according to the joint distribution~(\ref{eq:joint}), so $\langle s_i^{t'}|\by^{t'}\rangle=\tanh y_i^{t'}$.
Moreover, given $\by^{t'}$, the spins $\bs^{t'}$ and the future $\{\bs^{t''}\}_{t''>t'}$ are independent, because the future is generated from $\by^{t'}$ alone.
Hence $\langle s_i^t\tanh y_i^{t'}\rangle=\langle s_i^ts_i^{t'}\rangle$, and \mt{eq:response} becomes the discrete-time FDT, \mt{eq:FDT_discrete}, for any $N$.

\section{SyncMC update for the ROM}
\label{sec:rom}
For the ROM, $\lambda_{\min}(J)=-1$, so any $c\ge1$ is allowed.
Writing $J=2P-I$, where $P$ is the projection onto the $N/2$-dimensional subspace spanned by the $+1$ eigenvectors, and using $P^2=P$, \mt{eq:SyncMC_y} with $\bm h=0$ becomes
\begin{align}
\by^t&=\beta(c-1)\bs^t+\sqrt{\beta(c-1)}\,\bm z^t+2\beta P\bm{v}^t,\nonumber\\
\bm{v}^t&=\bs^t+\gamma\bm{z}^t,\qquad \gamma=\frac{\sqrt{c+1}-\sqrt{c-1}}{2\sqrt{\beta}} ,
\label{eq:ROM_SyncMC}
\end{align}
where the same Gaussian vector $\bm z^t$ appears in both places.
In words, apart from a local self-field and local noise, the fields are obtained by adding independent noise to the spins and projecting the result onto a random subspace of half the dimension.
For the minimal shift $c=1$ used in the main text, the local terms vanish, $\gamma=(2\beta)^{-1/2}$, and Eq.~\eqref{eq:ROM_SyncMC} reduces to $\by^t=2\beta P\bm v^t$; moreover $P=O_+O_+^{\top}$ with $O_+$ the $N\times N/2$ block of $O$, so this is the case $K=N/2$ of Eq.~\eqref{eq:SyncMC_eig}, and each step costs two products with $N\times N/2$ matrices.
Runs with $c>1$, which use the full matrix $O$ and the same coupling realizations, serve only to examine how the choice of $c$ affects the relaxation (Sec.~\ref{sec:c}).

\section{Derivation of the DMFT}
\label{sec:dmft}
We derive the DMFT for any coupling matrix of the form $J=O\Lambda O^{\top}$, with $O$ Haar distributed and the empirical distribution of the eigenvalues $\Lambda$ converging to a fixed distribution as $N\to\infty$.
Its R-transform $\RJ(z)$~\cite{GuionnetMaida2005} is written as a power series $\RJ(z)=\sum_{n\ge1}a_nz^n$; we assume $N^{-1}\Tr J\to0$ (a nonzero mean eigenvalue can be absorbed into $c$), so that $a_0=\RJ(0)=0$.
For the ROM, Eq.~(\ref{eq:RJ}) gives
\begin{equation}
\RJ(z)=\sum_{k\ge0}(-1)^k\,\frac{(2k)!}{k!\,(k+1)!}\,z^{2k+1}=z-z^3+2z^5-5z^7+\cdots,
\qquad \RJ(z)=z\,[1-\RJ(z)^2].
\label{eq:RJ_series}
\end{equation}
It is convenient to use the local field $\bu^t\equiv\by^t/\beta$.
With $K\equiv J+cI$, one SyncMC step~(\ref{eq:SyncMC}) reads
\begin{equation}
\bu^t=K\bs^t+\beta^{-1/2}K^{1/2}\bm z^t,\qquad
P(s_i^{t+1}\,|\,u_i^t)=\frac{e^{\beta s_i^{t+1}u_i^t}}{2\cosh\beta u_i^t},
\label{eq:u_update}
\end{equation}
i.e., given $\bs^t$, $\bu^t$ is Gaussian with mean $K\bs^t$ and covariance $K/\beta$.
A field $h_i^t$ in the main text enters as $u_i^t\to u_i^t+h_i^t$.

\subsection{Generating functional}
The probability of a path $\{\bs^t,\bu^t\}_{t=0}^{L}$ is $P(\bs^0)\prod_t P(\bs^{t+1}|\bu^t)P(\bu^t|\bs^t)$.
Writing the Gaussian factor as a Fourier integral over a response field $\hu^t$ (integrated along the imaginary axis)~\cite{Martin1973,DeDominicis1978},
\begin{equation}
P(\bu^t|\bs^t)\propto\int d\hu^t\,
\exp\!\Big[-\hu^{t\top}(\bu^t-K\bs^t)+\frac{1}{2\beta}\,\hu^{t\top}K\hu^t\Big],
\label{eq:MSR}
\end{equation}
the part of the path weight that depends on $J$ is
\begin{equation}
\exp\!\Big[\sum_t\Big(\hu^{t\top}J\bs^t+\frac{1}{2\beta}\hu^{t\top}J\hu^t\Big)\Big]
=\exp\!\Big[\frac12\Tr\big(J\,X\mathcal A X^{\top}\big)\Big],
\qquad
\mathcal A=\begin{pmatrix}0&I\\ I&\beta^{-1}I\end{pmatrix},
\label{eq:Jpart}
\end{equation}
where $X=(\bs^0,\dots,\bs^L,\hu^0,\dots,\hu^L)$ is an $N\times2(L+1)$ matrix.
The terms proportional to $c$ are local and are kept as they are.
Derivatives of the generating functional with respect to $h_i^t$ bring down $\hat u_i^t$, so that averages of $\hat u$ give response functions; in particular $\langle\hat u_i^{t}s_i^{\tau}\rangle=\partial\langle s_i^\tau\rangle/\partial h_i^t$.

\subsection{Average over the rotation}
The matrix $X\mathcal AX^{\top}$ has rank at most $2(L+1)$, which stays finite as $N\to\infty$.
For such low-rank arguments the average over $O$ is given by the asymptotics of spherical integrals~\cite{GuionnetMaida2005}:
\begin{equation}
\mathbb E_O\exp\!\Big[\frac12\Tr\big(JX\mathcal AX^{\top}\big)\Big]
\simeq\exp\!\big[N\,\Tr\,G(\mathcal A\mathcal B)\big],
\qquad G(z)=\frac12\int_0^z\RJ(w)\,dw,
\label{eq:GM}
\end{equation}
where $\mathcal B=X^{\top}X/N$ collects the order parameters,
\begin{equation}
\mathcal B=\begin{pmatrix}\tilde\C&\mathsf S^{\top}\\ \mathsf S&\mathsf W\end{pmatrix},\qquad
\tilde\C_{t\tau}=\frac1N\bs^t\!\cdot\bs^\tau,\quad
\mathsf S_{t\tau}=\frac1N\hu^t\!\cdot\bs^\tau,\quad
\mathsf W_{t\tau}=\frac1N\hu^t\!\cdot\hu^\tau .
\end{equation}
(For a rank-one argument, Eq.~\eqref{eq:GM} reduces to $\mathbb E_O\exp[\frac N2\theta\,\bm e^{\top}J\bm e]\simeq\exp[\frac N2\int_0^\theta\RJ]$ for a unit vector $\bm e$, which defines the R-transform.)
The order parameters are introduced with delta functions and conjugate variables in the usual way, and the integrals are evaluated by the saddle-point method for $N\to\infty$.

\subsection{Saddle point and the effective single-site process}
At the saddle point, causality gives $\mathsf W=0$ (the normalization of the path probability), $\tilde\C=\C$, the spin correlation matrix, and $\mathsf S=\R^{\top}$, where $\R_{t\tau}=R(t,\tau)$ is the response matrix of the main text (strictly lower triangular).
Then
\begin{equation}
\mathcal A\mathcal B=\begin{pmatrix}\R^{\top}&0\\ \C+\beta^{-1}\R^{\top}&\R\end{pmatrix}
\label{eq:AB}
\end{equation}
is block triangular.
Varying $N\Tr G(\mathcal A\mathcal B)$ with respect to $\mathcal B$ gives $d\,\Tr G(\mathcal A\mathcal B)=\Tr[\mathcal F\mathcal A\,d\mathcal B]$ with $\mathcal F=G'(\mathcal A\mathcal B)=\frac12\RJ(\mathcal A\mathcal B)$, and the conjugate variables are fixed by $\mathcal F$.
For a block-triangular argument,
\begin{equation}
f\!\begin{pmatrix}X&0\\ P&Y\end{pmatrix}=\begin{pmatrix}f(X)&0\\ f^{[1]}_{Y,X}[P]&f(Y)\end{pmatrix},
\qquad
f^{[1]}_{Y,X}[P]\equiv\sum_{n}f_n\sum_{k=0}^{n-1}Y^{k}PX^{n-1-k},
\label{eq:divdiff}
\end{equation}
for $f(z)=\sum_nf_nz^n$; $f^{[1]}$ is the matrix form of the divided difference, reducing to $P\,[f(Y)-f(X)]/(Y-X)$ when $X$, $Y$, and $P$ are numbers~\cite{Higham2008}.
Collecting the terms of the saddle-point action that involve site $i$, all sites decouple, and each follows the effective single-site process [\mt{eq:DMFT}]
\begin{subequations}
\label{eq:DMFT_sigma}
\begin{align}
u^t&=c\,s^t+\sum_{\tau<t}\U_{t\tau}\,s^\tau+\eta^t,\qquad
s^{t+1}=\pm1\ \text{with probability}\ \frac{1\pm\tanh\beta u^t}{2},
\label{eq:DMFT_sigma_a}\\
\U&=\RJ(\R),
\label{eq:DMFT_sigma_U}\\
\Nn&\equiv\langle\bm\eta\bm\eta^{\top}\rangle
=\RJ^{[1]}{}_{\!\R,\R^{\top}}[\C]
+\beta^{-1}\,(z\RJ)^{[1]}{}_{\!\R,\R^{\top}}[I]
+\frac{c}{\beta}\,I ,
\label{eq:DMFT_sigma_N}
\end{align}
\end{subequations}
where $\eta^t$ is a Gaussian process with zero mean, and $(z\RJ)^{[1]}$ denotes Eq.~\eqref{eq:divdiff} for the function $z\RJ(z)$.
Explicitly,
\begin{equation}
\Nn=\sum_{n\ge1}a_n\sum_{k=0}^{n-1}\R^k\,\C\,(\R^{\top})^{n-1-k}
+\frac1\beta\sum_{n\ge1}a_n\sum_{k=0}^{n}\R^k(\R^{\top})^{n-k}+\frac c\beta I .
\label{eq:N_series}
\end{equation}
Here we used that the saddle point produces the covariance $2\,\mathrm{sym}[\mathcal F_{21}+\beta^{-1}\mathcal F_{22}]+(c/\beta)I$, where $\mathrm{sym}[M]=(M+M^\top)/2$; the terms $\pm\frac1{2\beta}[\RJ(\R)+\RJ(\R)^{\top}]$ that arise from the $\beta^{-1}\R^{\top}$ block of Eq.~\eqref{eq:AB} and from $\mathcal F_{22}$ cancel, which leaves the symmetric form~\eqref{eq:N_series}.
Both $\C$ and $\R$ are determined self-consistently by the process~\eqref{eq:DMFT_sigma}:
\begin{equation}
\C_{t\tau}=\langle s^ts^\tau\rangle,\qquad
\R_{t\tau}=\frac{\partial\langle s^t\rangle}{\partial h^\tau}=\beta\big\langle s^t\big(s^{\tau+1}-\tanh\beta u^\tau\big)\big\rangle\quad(\tau<t),
\label{eq:selfcons}
\end{equation}
where the second form is the estimator of \mt{eq:response} applied to the single-site process (equivalently, $\R$ can be obtained from Novikov's formula~\cite{Novikov1965}, $\langle s^t\eta^u\rangle=\sum_\tau\R_{t\tau}\Nn_{\tau u}$).
Equations~\eqref{eq:DMFT_sigma}--\eqref{eq:selfcons} are exact for $N\to\infty$ and hold for any initial distribution of $\bs^0$ that is independent of $O$.

The structure of Eq.~\eqref{eq:DMFT_sigma} is simple: the average over the rotation replaces the coupling term $J\bs^t$ by a memory term $\RJ(\R)$ acting on the past spins of the same site (the Onsager reaction, retarded in time) and a Gaussian noise $\eta^t$ whose covariance depends on $\C$ and $\R$.
For the SK model, $\RJ(z)=z$, and Eq.~\eqref{eq:DMFT_sigma} reduces to $\U=\R$ and $\Nn=\C+\beta^{-1}(\R+\R^{\top})+(c/\beta)I$.

\subsection{Evaluation of the kernels}
Because $\R$ is strictly lower triangular, $\R^{L+1}=0$, and all power series above are finite.
They need not be summed term by term.
For the ROM, Eq.~\eqref{eq:RJ_series} gives the Dyson-type relation
\begin{equation}
\U=\R\,(I-\U^2),
\label{eq:Dyson}
\end{equation}
and, applying the same relation to the block matrix~\eqref{eq:divdiff}, the divided difference $\mathsf L[P]\equiv\RJ^{[1]}{}_{\R,\R^{\top}}[P]$ obeys
\begin{equation}
\mathsf L[P]=P-P\,(\U^{\top})^2-\R\,\big(\mathsf L[P]\,\U^{\top}+\U\,\mathsf L[P]\big).
\label{eq:Sylvester}
\end{equation}
With these, Eq.~\eqref{eq:DMFT_sigma_N} becomes
\begin{equation}
\Nn=\mathsf L[\C]+\beta^{-1}\big(\U^{\top}+\R\,\mathsf L[I]\big)+\frac c\beta I .
\label{eq:N_rec}
\end{equation}
Equations~\eqref{eq:Dyson}--\eqref{eq:N_rec} can be solved row by row in time.
We checked them against the direct sums~\eqref{eq:N_series} for random strictly lower-triangular $\R$ and positive-definite $\C$ (agreement to machine precision).

Every quantity at time $t$ depends only on earlier times, so the DMFT is solved in a single pass forward in time.
At step $t$, $s^t$ has already been sampled for a large number of independent trajectories; row $t$ of $\C$ and $\R$ is estimated from them with Eq.~\eqref{eq:selfcons}; row $t$ of $\U$ and $\Nn$ follows from Eqs.~\eqref{eq:Dyson}--\eqref{eq:N_rec}; $\eta^t$ is drawn conditionally on $\eta^0,\dots,\eta^{t-1}$ (one new row of the Cholesky factor of $\Nn$); and $s^{t+1}$ is sampled.
To reduce the statistical error, we replace $s^t$ by its conditional mean $\tanh\beta u^{t-1}$ in the estimators whenever this does not change the expectation, e.g., $\C_{t\tau}=\langle\tanh(\beta u^{t-1})s^\tau\rangle$ for $\tau<t$ and $\R_{t,t-1}=\beta\langle1-\tanh^2\beta u^{t-1}\rangle$.

\subsection{An equivalent representation for the ROM}
For the ROM, the average over $O$ can also be applied only to the projection term $2\beta P\bm v^t$ of Eq.~\eqref{eq:ROM_SyncMC}~\cite{Opper2016}.
The memory term then acts on the noisy variable $v^t=s^t+\gamma z^t$ rather than on the spins,
\begin{equation}
y^t=\beta(c-1)s^t+\sqrt{\beta(c-1)}\,z^t+\beta v^t+\beta\sum_{\tau<t}\big[\RJ(\R)\big]_{t\tau}v^\tau+2\beta\phi^t,
\label{eq:DMFT_v}
\end{equation}
where $z^t$ is a standard normal variable drawn afresh at each step and $\phi^t$ is Gaussian with covariance
\begin{equation}
\mathsf\Sigma=\frac{1}{4}\sum_{k\ge2}a_{k-1}\sum_{l=0}^{k-2}\R^{l}\,\C_v\,(\R^{\top})^{k-2-l},
\qquad
\C_v=\C+\gamma^2I+\mathsf Y+\mathsf Y^{\top},\quad
\mathsf Y=\gamma^2\R\,[I+\RJ(\R)]+\gamma\sqrt{(c-1)/\beta}\,\R .
\label{eq:Sigma_v}
\end{equation}
The correlation matrix $\C_v$ of $v$ differs from $\C$ because $z^\tau$ affects the later spins through $y^\tau$; it follows from integrating by parts over $z^\tau$.
With $\U=\RJ(\R)$, the covariance obeys $\mathsf\Sigma=\frac{1}{4}\C_v-\R(\U\mathsf\Sigma+\mathsf\Sigma\U^{\top})-\frac{1}{4}\C_v(\U^{\top})^2$, which can be solved row by row.
Equation~\eqref{eq:DMFT_sigma} is simpler, since it needs neither $v$ nor $\C_v$, and it applies to any rotation-invariant $J$.
The DMFT curves in Figs.~\ref{fig:fig2} and \ref{fig:fig3} of the main text were computed with Eq.~\eqref{eq:DMFT_v}; the two representations agree within the sampling error (Sec.~\ref{sec:check}).

\section{Stationary solution above $T_{\rm d}$}
\label{sec:tti}
\subsection{Stationary equations}
If the initial configuration is drawn from the canonical distribution, SyncMC is a stationary process, and Sec.~\ref{sec:response} shows that it satisfies the FDT, \mt{eq:FDT_discrete}, for any $N$.
The DMFT describes the same process for $N\to\infty$; its stationary solution therefore depends only on time differences, $C(t,t')=C(t-t')$ and $R(t,t')=R(t-t')$, and satisfies
\begin{equation}
R(\tau)=\beta\,[C(\tau-1)-C(\tau)]\qquad(\tau\ge1).
\label{eq:FDT_tau}
\end{equation}
With $\hat f(\omega)=\sum_\tau f(\tau)e^{i\omega\tau}$ and $\hat C_+(\omega)=\sum_{\tau\ge0}C(\tau)e^{i\omega\tau}$, Eq.~\eqref{eq:FDT_tau} reads
\begin{equation}
\hat R(\omega)=\beta\,\big[1-(1-e^{i\omega})\,\hat C_+(\omega)\big].
\label{eq:FDT_omega}
\end{equation}
For time-translation-invariant kernels, products of $\R$ and $\R^{\top}$ become products of $\hat R(\omega)$ and $\hat R(\omega)^*$, and the divided differences~\eqref{eq:divdiff} become ordinary ones.
The memory kernel and the noise spectrum of Eq.~\eqref{eq:DMFT_sigma} are then
\begin{align}
\hat U(\omega)&=\RJ\big(\hat R(\omega)\big),\nonumber\\
\hat N(\omega)&=\hat C(\omega)\,\frac{\RJ(\hat R)-\RJ(\hat R^*)}{\hat R-\hat R^*}
+\frac1\beta\,\frac{\hat R\,\RJ(\hat R)-\hat R^*\RJ(\hat R^*)}{\hat R-\hat R^*}+\frac c\beta ,
\label{eq:TTI_kernels}
\end{align}
with $\hat C(\omega)=\hat C_+(\omega)+\hat C_+(\omega)^*-1$.
Together with Eq.~\eqref{eq:FDT_omega}, these express $\hat U$ and $\hat N$ through $C$ alone, and the correlation of the stationary single-site process closes the equations:
\begin{equation}
C(\tau)=\langle s^{t+\tau}s^t\rangle\equiv\mathcal M_T[C](\tau).
\label{eq:MT}
\end{equation}

\subsection{Dynamical transition}
\label{sec:Td}
As $T$ decreases toward $T_{\rm d}$, $C(\tau)$ develops a plateau: it first decays to a value $q$ on a short time scale, stays there for a long time, and only then decays to zero (the $\alpha$ relaxation) on a time scale $\tau_\alpha$ that diverges as $T\to T_{\rm d}^+$.
On the plateau, the long-lived part of the local field is effectively frozen.
The analysis of Sec.~\ref{sec:aging}, with the fluctuation-dissipation ratio set to $x=1$ as appropriate for an equilibrium process, shows that the plateau height $q$ must satisfy
\begin{equation}
q=\mathcal M_T(q;1),\qquad
\mathcal M_T(q;x)\equiv\frac{\int\!Dy\,\cosh^{x}(\kappa y)\tanh^2(\kappa y)}{\int\!Dy\,\cosh^{x}(\kappa y)},\qquad
\kappa^2=\frac{\beta}{x}\Big[\RJ\big(\beta(1-q)+x\beta q\big)-\RJ\big(\beta(1-q)\big)\Big],
\label{eq:plateau}
\end{equation}
where $Dy$ is the standard Gaussian measure.
For $x=1$, the shift $y\to y+\kappa$ gives the equivalent form $q=\int Dy\tanh^2(\kappa y+\kappa^2)$ with $\kappa^2=\beta[\RJ(\beta)-\RJ(\beta(1-q))]$.
Equation~\eqref{eq:plateau} is the 1RSB equation for the overlap $q$ with Parisi parameter $m=x$~\cite{Marinari1994,Parisi1995,Cherrier2003}; for $x=1$ it is the non-ergodicity equation of the dynamics.
Above $T_{\rm d}$ it has only the solution $q=0$; $T_{\rm d}$ is the highest temperature at which a nonzero solution appears, where the solution is marginal:
\begin{equation}
\frac{\partial\mathcal M_T(q;1)}{\partial q}=1
\quad\Longleftrightarrow\quad
\beta^2\RJ'\big(\beta(1-q)\big)\,\big\langle(1-m^2)^2\big\rangle_1=1,
\label{eq:Td}
\end{equation}
where $\langle\cdots\rangle_x$ is the average with the weight in Eq.~\eqref{eq:plateau} and $m=\tanh\kappa y$; the two forms coincide at $x=1$ (checked numerically).
For the ROM we obtain
\begin{equation}
\beta_{\rm d}=7.483,\qquad T_{\rm d}=0.1336,\qquad q_{\rm d}=0.961 .
\label{eq:Td_values}
\end{equation}

\section{Aging below $T_{\rm d}$}
\label{sec:aging}
Below $T_{\rm d}$, the dynamics started from a random configuration does not become stationary.
We analyze Eq.~\eqref{eq:DMFT_sigma} for $t'\to\infty$ under the three time scales stated in the main text:
(i) $t-t'=O(1)$, (ii) $1\ll t-t'\ll t'$, and (iii) $t-t'=O(t')$.
This is the weak ergodicity breaking scenario of Cugliandolo and Kurchan~\cite{Cugliandolo1993,Cugliandolo2003}: we assume that $C(t,t')\to0$ as $t/t'\to\infty$ at every $T<T_{\rm d}$, i.e., that the system does not retain a memory of its initial configuration, so that the results below apply wherever this assumption holds.
Correspondingly, we write
\begin{equation}
C(t,t')=C_{\rm st}(t-t')+C_{\rm ag}(t,t'),\qquad R(t,t')=R_{\rm st}(t-t')+R_{\rm ag}(t,t'),
\end{equation}
where $C_{\rm st}(\tau)$ decays from $1-q$ to $0$ on scale (i) and $C_{\rm ag}$ decays from $q$ to $0$ on scale (iii); here $q=q_{\rm EA}$ is the plateau.
On scale (i), the FDT holds, $R_{\rm st}(\tau)=\beta[C_{\rm st}(\tau-1)-C_{\rm st}(\tau)]$, so that
\begin{equation}
\bar\chi\equiv\sum_{\tau\ge1}R_{\rm st}(\tau)=\beta(1-q).
\label{eq:chibar}
\end{equation}
The memory kernel and the noise inherit the same split, and the local field of Eq.~\eqref{eq:DMFT_sigma_a} separates into three parts,
\begin{equation}
u^t=\underbrace{c\,s^t+\eta^{(c),t}}_{\text{local}}
+\underbrace{\sum_{\tau<t}U_{\rm st}(t-\tau)s^\tau+\eta^{{\rm st},t}}_{\text{scale (i)}}
+\underbrace{\sum_{\tau<t}U_{\rm ag}(t,\tau)s^\tau+\eta^{{\rm ag},t}}_{\text{scale (iii)}\ \equiv\ z^t},
\label{eq:three_levels}
\end{equation}
where $\eta^{(c)}$ is the part of the noise with covariance $(c/\beta)I$.

\subsection{Scales (i) and (ii): equilibrium within a state}
On scales (i) and (ii), the slow field $z^t$ is constant.
The local part is the SyncMC update of a single spin with self-coupling $c$; by itself it satisfies detailed balance with respect to $P(s)\propto e^{\beta\tilde hs}$, where $\tilde h$ is the rest of the field, and therefore does not affect the stationary distribution (it affects only the speed of the dynamics; this is why the long-time properties do not depend on $c$).
The part on scale (i) is the stationary process of Sec.~\ref{sec:tti} with $C$ replaced by $C_{\rm st}$; it satisfies the FDT, and its memory kernel has total weight
\begin{equation}
\sum_{\tau\ge1}U_{\rm st}(\tau)=\RJ(\bar\chi).
\label{eq:Onsager}
\end{equation}
This memory term, together with the local term $c\,s^t$, is the reaction of the field to the spin itself (the Onsager reaction): averaged over scale (i), it adds $[c+\RJ(\bar\chi)]\,m$ to the field, where $m$ is the magnetization averaged over scale (i).
It does not, however, act on the spin as an additional external field; it cancels against the fast noise, as we now show.

On scales (i) and (ii) the spin is in equilibrium within a metastable state.
In equilibrium, the joint distribution of the spins and the local fields $\bu=\by/\beta$ follows from \mt{eq:joint}:
\begin{equation}
P(\bs,\bu)\propto\exp\!\Big[-\frac\beta2\,\bu^{\top}K^{-1}\bu+\beta\sum_ju_js_j\Big].
\label{eq:joint_u}
\end{equation}
Given $\bu$, each spin depends only on its own field, $P(s_i|\bu)=e^{\beta s_iu_i}/2\cosh\beta u_i$.
Summing over all spins and integrating over all fields except those of site $i$ gives, exactly,
\begin{equation}
P(s_i,u_i)\propto P_{\rm cav}(u_i)\,e^{\beta s_iu_i},
\label{eq:cavity}
\end{equation}
where $P_{\rm cav}(u_i)$ is obtained from Eq.~\eqref{eq:joint_u} with the factor $e^{\beta s_iu_i}$ removed, i.e., it is the distribution of the field on site $i$ when site $i$ does not act back on the others.
Within a state, $P_{\rm cav}$ is Gaussian for a mean-field model, with the slow field $z$ as its mean and some variance $V$ from the fast fluctuations.
Equation~\eqref{eq:cavity} then gives
\begin{equation}
P(u_i\,|\,s_i)=\mathcal N\big(u_i;\,z+\beta V s_i,\,V\big),\qquad
P(s_i)\propto\int du\,\mathcal N(u;z,V)\,e^{\beta s_iu}=e^{\beta zs_i}\,e^{\beta^2V/2}.
\label{eq:cancel}
\end{equation}
The first relation shows that the spin shifts the mean of its own field by $\beta V s_i$; this shift is the Onsager reaction, so that $\beta V=c+\RJ(\bar\chi)$.
The second relation shows that this shift and the Gaussian fluctuations of the field combine into the factor $e^{\beta^2V/2}$, which does not depend on $s_i$: the reaction and the fast noise cancel in the distribution of the spin.
Hence the magnetization averaged over scales (i) and (ii) is determined by the slow field alone,
\begin{equation}
m=\tanh\beta z ,
\label{eq:m_tanh}
\end{equation}
independently of $c$ and $V$; $z$ plays the role of the cavity field of the state~\cite{Parisi1995}, and $q=\langle m^2\rangle$.
Two checks: (a) Eq.~\eqref{eq:m_tanh} gives $\langle\partial m/\partial z\rangle=\beta(1-q)=\bar\chi$, in agreement with Eq.~\eqref{eq:chibar}; (b) above $T_{\rm d}$ ($z=0$, $q=0$, $\bar\chi=\beta$) the first relation of Eq.~\eqref{eq:cancel} gives $\langle u_is_i\rangle=\beta V=c+\RJ(\beta)$, which is exact, since $N^{-1}\sum_i\langle u_is_i\rangle=N^{-1}\langle\bs^{\top}K\bs\rangle=c-2e$ with $e=-\RJ(\beta)/2$.

\subsection{Scale (iii): aging}
On scale (iii) we use the time variable $\lambda=\ln h(t)$, where $h(t)$ is the (unknown) effective age, $h(t)=t$ for simple aging, and assume that the aging parts depend only on $\ell=\lambda-\lambda'$~\cite{Cugliandolo1993,Cugliandolo2003}:
\begin{equation}
C_{\rm ag}(t,t')=c(\ell),\qquad R_{\rm ag}(t,t')\,dt'=r(\ell)\,d\lambda',\qquad c(0)=q,\ c(\infty)=0 .
\end{equation}
On this scale the stationary response acts instantaneously, $R_{\rm st}(t-t')\,dt'\to\bar\chi\,\delta(\lambda-\lambda')\,d\lambda'$, and the $\beta^{-1}$ terms of Eq.~\eqref{eq:N_series} are smaller by a factor $1/t$.
With $\hat f(\nu)=\int d\ell\,f(\ell)\,e^{-i\nu\ell}$ (a Fourier transform in $\lambda$, i.e., a Mellin transform in $h$), the block matrix~\eqref{eq:AB} restricted to scale (iii) has the diagonal entries $\bar\chi+\hat r(\nu)$ and $\bar\chi+\hat r(\nu)^*$ and the off-diagonal entry $\hat c(\nu)$.
Subtracting the fast parts, the aging memory kernel and noise spectrum are
\begin{equation}
\hat U_{\rm ag}(\nu)=\RJ\big(\bar\chi+\hat r(\nu)\big)-\RJ(\bar\chi),\qquad
\hat N_{\rm ag}(\nu)=\hat c(\nu)\,\frac{\RJ\big(\bar\chi+\hat r(\nu)\big)-\RJ\big(\bar\chi+\hat r(\nu)^*\big)}{\hat r(\nu)-\hat r(\nu)^*} .
\label{eq:ag_kernels}
\end{equation}
Now assume that the response and correlation on scale (iii) are related by a constant ratio $x$ [\mt{eq:GFDT_diff}],
\begin{equation}
r(\ell)=-x\beta\,c'(\ell)\qquad(\ell>0).
\label{eq:ag_FDT}
\end{equation}
Then $\hat c(\nu)=i[\hat r(\nu)-\hat r(\nu)^*]/(x\beta\nu)$, and Eq.~\eqref{eq:ag_kernels} gives the same relation for the kernels,
\begin{equation}
\hat N_{\rm ag}(\nu)=\frac{i\,[\hat U_{\rm ag}(\nu)-\hat U_{\rm ag}(\nu)^*]}{x\beta\nu}
\quad\Longleftrightarrow\quad
U_{\rm ag}(\ell)=-x\beta\,N_{\rm ag}'(\ell)\quad(\ell>0).
\label{eq:kernel_FDT}
\end{equation}
Thus the slow field is driven by a memory and a noise that are related as in equilibrium, but at the inverse temperature $x\beta$.
Integrating Eq.~\eqref{eq:kernel_FDT} over $\ell>0$ and using $\hat r(0)=x\beta q$ gives the equal-time variance of the slow noise,
\begin{equation}
\Delta\equiv N_{\rm ag}(0)=\frac{\hat U_{\rm ag}(0)}{x\beta}
=\frac{\RJ(\bar\chi+x\beta q)-\RJ(\bar\chi)}{x\beta}.
\label{eq:Delta}
\end{equation}
The slow field obeys
\begin{equation}
z(\lambda)=\int_{-\infty}^{\lambda}d\lambda'\,U_{\rm ag}(\lambda-\lambda')\tanh\beta z(\lambda')+\eta^{\rm ag}(\lambda),\qquad
\langle\eta^{\rm ag}(\lambda)\eta^{\rm ag}(\lambda')\rangle=N_{\rm ag}(\lambda-\lambda'),
\label{eq:slow}
\end{equation}
where the memory acts on the magnetization~\eqref{eq:m_tanh} because $U_{\rm ag}$ varies only on scale (iii).

\subsection{Stationary distribution of the slow field and the plateau}
For an exponential kernel, $N_{\rm ag}(\ell)=\Delta e^{-|\ell|}$, Eqs.~\eqref{eq:kernel_FDT} and \eqref{eq:slow} are equivalent to the Markov process $dz=[-z+x\beta\Delta\tanh\beta z]\,d\lambda+\sqrt{2\Delta}\,dW$, whose Fokker--Planck equation has the stationary solution
\begin{equation}
P_{\rm st}(z)\propto\cosh^{x}(\beta z)\,e^{-z^2/(2\Delta)} .
\label{eq:Pst}
\end{equation}
We verified numerically that Eq.~\eqref{eq:Pst} does not depend on the shape of the kernel as long as Eq.~\eqref{eq:kernel_FDT} holds (Sec.~\ref{sec:slowcheck}).
The plateau is $q=\int dz\,P_{\rm st}(z)\tanh^2\beta z$.
With $\bar\chi=\beta(1-q)$ from Eq.~\eqref{eq:chibar} and $\kappa=\beta\sqrt\Delta$, this is Eq.~\eqref{eq:plateau}:
the aging dynamics reproduces the 1RSB equation with the Parisi parameter $m$ replaced by the fluctuation-dissipation ratio $x$.
For $x=1$ the slow process is an equilibrium process, and Eq.~\eqref{eq:plateau} becomes the plateau condition used in Sec.~\ref{sec:Td}.

\subsection{Condition fixing $x$}
Equation~\eqref{eq:plateau} alone determines $q$ as a function of $x$; a second condition is needed.
As for the spherical $p$-spin model~\cite{Cugliandolo1993}, it comes from matching scales (i) and (ii): for the correlation to approach the plateau and leave it only on scale (iii), the plateau must be marginally stable with respect to the fast dynamics.
Linearizing the fast equations about the plateau with the slow field held fixed gives the stability parameter $1-\beta^2\RJ'(\bar\chi)\langle(1-m^2)^2\rangle_x$, which must vanish,
\begin{equation}
\beta^2\RJ'\big(\beta(1-q)\big)\,\big\langle(1-m^2)^2\big\rangle_x=1 .
\label{eq:marginal}
\end{equation}
This is the marginal stability (replicon) condition of the 1RSB solution with $m=x$, and for $x=1$ it reduces to the condition~\eqref{eq:Td} for $T_{\rm d}$.
Solving Eqs.~\eqref{eq:plateau} and \eqref{eq:marginal} gives $q_{\rm EA}(T)$ and $x(T)$ (Table~\ref{tab:threshold}); at $T=0.1$, $q_{\rm EA}=0.9815$ and $x=0.679$.

\begin{table}[t]
\caption{Plateau height $q_{\rm EA}$ and fluctuation-dissipation ratio $x$ on scale (iii) from Eqs.~\eqref{eq:plateau} and \eqref{eq:marginal}, and the width $\kappa$ of the slow field. At $T_{\rm d}$, $x=1$.}
\label{tab:threshold}
\begin{ruledtabular}
\begin{tabular}{cccc}
$T$ & $q_{\rm EA}$ & $x$ & $\kappa$ \\
\hline
$0.1336\ (T_{\rm d})$ & 0.9611 & 1.000 & 2.232 \\
0.13 & 0.9640 & 0.956 & 2.332 \\
0.12 & 0.9710 & 0.850 & 2.626 \\
0.11 & 0.9767 & 0.759 & 2.952 \\
0.10 & 0.9815 & 0.679 & 3.325 \\
0.09 & 0.9855 & 0.606 & 3.762 \\
0.08 & 0.9889 & 0.539 & 4.289 \\
0.07 & 0.9917 & 0.475 & 4.944 \\
0.06 & 0.9940 & 0.414 & 5.791 \\
0.05 & 0.9959 & 0.354 & 6.940 \\
\end{tabular}
\end{ruledtabular}
\end{table}

\subsection{Generalized FDT}
Combining the FDT on scale (i) with Eq.~\eqref{eq:ag_FDT} on scale (iii) gives, for large $t'$,
\begin{equation}
R(t,t')=X\big(C(t,t')\big)\,\beta\,[C(t,t'+1)-C(t,t')],\qquad
X(C)=\begin{cases}1,&C>q_{\rm EA},\\ x,&C<q_{\rm EA},\end{cases}
\end{equation}
and summing over $t'$ gives the integrated response $\chi(t,t_w)=\beta(1-q_{\rm EA})+x\beta\,[q_{\rm EA}-C(t,t_w)]$ for $C<q_{\rm EA}$, which is \mt{eq:GFDT}.

\section{Additional numerical results}
\label{sec:num}

\subsection{Two forms of the DMFT}
\label{sec:check}
We solved Eqs.~\eqref{eq:DMFT_sigma} and \eqref{eq:DMFT_v} for a quench from a random configuration, each with two independent runs of $2^{18}$ trajectories, up to $t=32$.
The largest differences over $0\le t'<t\le32$ between the averages of the two forms are $|\Delta C|=0.0048$ and $T|\Delta\chi|=0.0033$ ($T=0.3$, $c=1$), $0.0023$ and $0.0010$ ($T=0.1$, $c=1$), and $0.0045$ and $0.0028$ ($T=0.3$, $c=1.2$); they are smaller than the difference between two runs of the same form ($0.003$--$0.009$ in $C$).
Both forms also agree with direct SyncMC simulations ($N=3000$) at early times.

\subsection{Stationary distribution of the slow field}
\label{sec:slowcheck}
To check that Eq.~\eqref{eq:Pst} holds beyond the exponential kernel, we simulated Eq.~\eqref{eq:slow} directly for $T=0.1$, $x=0.679$, and $\Delta=0.1105$ (Table~\ref{tab:threshold}), with $U_{\rm ag}$ obtained from Eq.~\eqref{eq:kernel_FDT}, for three kernel shapes, and measured $\langle\tanh^2\beta z\rangle$.
Extrapolating the step size $d\lambda\to0$, we obtain $0.9815$ for $N_{\rm ag}(\ell)=\Delta e^{-|\ell|}$ and for $\Delta e^{-\ell^2/2}$, and $0.981$ for $\Delta/(1+|\ell|)^2$ (memory truncated at $\ell=12$), compared with $0.9815$ from Eq.~\eqref{eq:Pst}.

\subsection{Dependence on the diagonal shift $c$}
\label{sec:c}
\begin{figure}[t]
\centering
\includegraphics[width=0.38\linewidth]{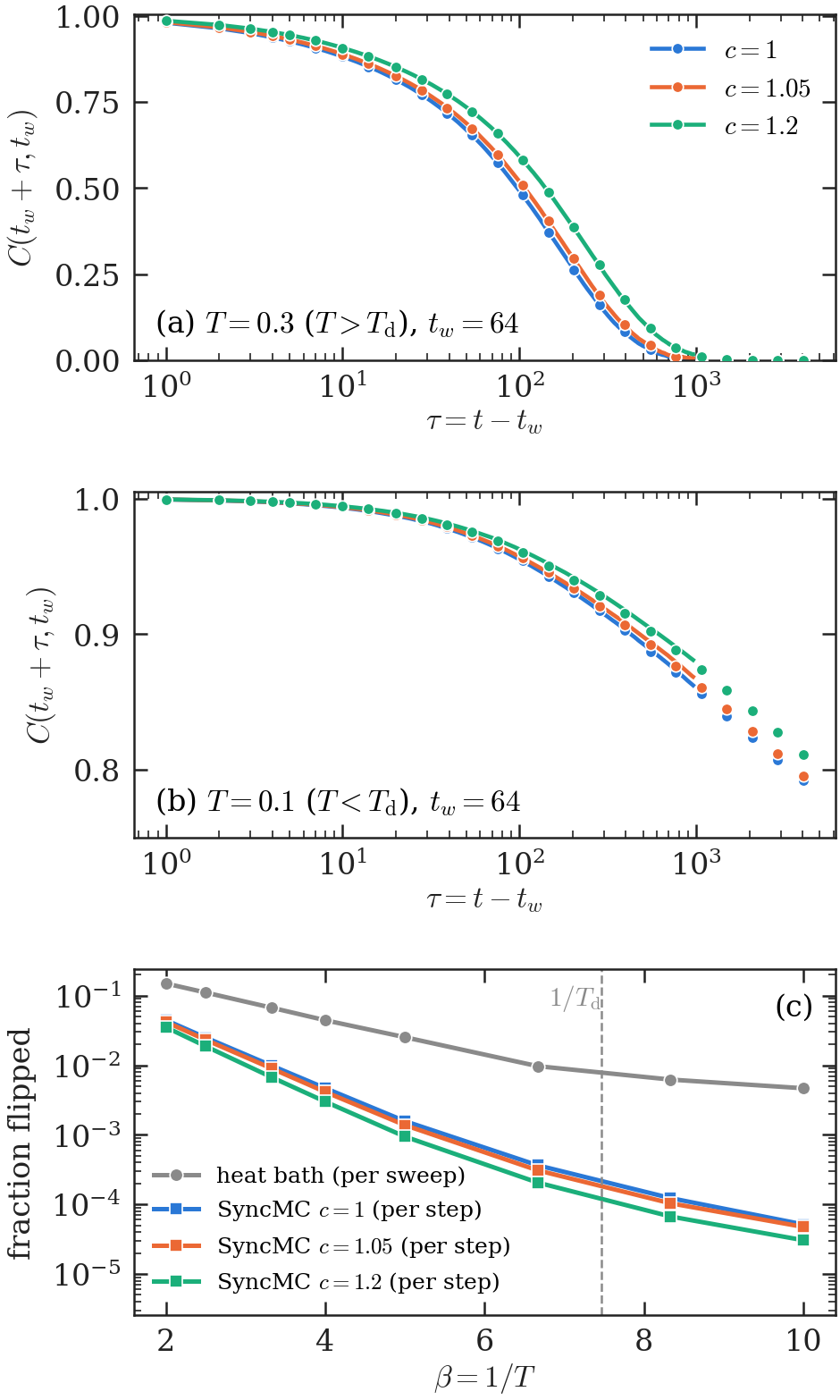}
\caption{Effect of the diagonal shift $c$ on the SyncMC dynamics.
(a), (b) $C(t_w+\tau,t_w)$ after a quench from a random configuration, for $t_w=64$, at $T=0.3$ and $0.1$; $c=1$, $1.05$, and $1.2$ with the same coupling realizations.
Symbols: SyncMC for $N=16384$, averaged over $4$ samples of $O$ with $16$ chains each; lines: DMFT, averaged over $4$ runs with $2^{19}$ trajectories each.
(c) Fraction of spins flipped per SyncMC step ($N=4096$, $16$ chains, $200$ steps after $1000$ steps of equilibration) and per sweep of sequential heat-bath updates ($N=2048$, $50$ sweeps after $300$), in equilibrium, versus $\beta$.}
\label{fig:c}
\end{figure}
Every $c\ge-\lambda_{\min}(J)$ gives the same stationary distribution but not the same dynamics.
Figure~\ref{fig:c} shows that a larger $c$ slows the relaxation, as expected from the self-field $\beta c\,s_i^t$ that favors keeping the current sign.
At $T=0.3$, the time at which $C(t_w+\tau,t_w)$ falls below $1/e$ is $\tau_{\rm rel}=147$, $163$, and $215$ for $c=1$, $1.05$, and $1.2$, i.e., $11\%$ and $46\%$ longer than for the minimal shift.
For the same reason, fewer spins flip per SyncMC step than per sequential sweep [Fig.~\ref{fig:c}(c)]: for $c=1$ the fraction is $4.5\times10^{-2}$ at $T=0.5$ and $5.2\times10^{-5}$ at $T=0.1$, compared with $0.15$ and $4.7\times10^{-3}$ for the heat bath (ratios $3.4$ and $91$), and it decreases further with $c$ (by $6\%$ and $21\%$ for $c=1.05$ and $1.2$ at $T=0.5$; $8\%$ and $41\%$ at $T=0.1$, where the statistical error is a few percent).
This ratio grows at low $T$ but does not depend on $N$, so the gain in span, $N/\log N$, wins for large systems.

\subsection{Long runs at finite $N$}
The DMFT can be solved only for $t\lesssim10^3$.
To look at longer times, we ran SyncMC with $c=1$ for $N=1000$--$4000$ up to $t=3.2\times10^6$, with $64$ samples of $O$ (one chain each), and measured the integrated response in two ways: with a random-sign field of strength $\epsilon=0.01$ switched on at $t_w$ in a copy driven by the same random numbers, and with the field-free estimator of \mt{eq:response}.
Fitting the generalized FDT, \mt{eq:GFDT}, to all data with $C<q_{\rm EA}$ (Table~\ref{tab:threshold}) gives the slopes in Table~\ref{tab:longruns} (errors from a jackknife over the samples).
They are consistent with Table~\ref{tab:threshold} within the errors; a trend with $t_w$ is not resolved, and for the largest $t_w$ the slope is poorly determined because $C$ decays only to $0.87$ ($T=0.12$) and $0.93$ ($T=0.10$) within the run.
For $\epsilon=0.02$ the field response overestimates $x$ at $T=0.12$ ($1.20\pm0.14$ at $t_w=10^5$, against $0.83\pm0.17$ for $\epsilon=0.01$ and $0.84\pm0.19$ from \mt{eq:response} on the same trajectories), whereas at $T=0.12$, $\epsilon=0.01$ agrees with the field-free estimate.
The plateau estimated from $C$ at $t-t_w\in[t_w/20,t_w/2]$ ($0.9654\pm0.0020$ at $T=0.1$, $t_w=1.6\times10^6$) lies partly in the aging regime and is therefore a lower bound on $q_{\rm EA}$.
\begin{table}[t]
\caption{Fluctuation-dissipation ratio $x$ from fits of the generalized FDT to SyncMC data for $N=4000$, $c=1$, and $64$ samples of $O$, from the response to a field $\epsilon=0.01$ and, in parentheses, from the field-free estimator of \mt{eq:response}; the last column is the prediction of Table~\ref{tab:threshold}.}
\label{tab:longruns}
\begin{ruledtabular}
\begin{tabular}{ccccc}
$T$ & $t_w=10^5$ & $t_w=4\times10^5$ & $t_w=1.6\times10^6$ & Table~\ref{tab:threshold} \\
\hline
0.12 & $0.79\pm0.09$ ($0.88\pm0.08$) & $0.89\pm0.11$ ($0.90\pm0.12$) & $0.82\pm0.19$ ($1.23\pm0.22$) & 0.850 \\
0.10 & $0.68\pm0.10$ ($0.65\pm0.07$) & $1.01\pm0.13$ ($0.63\pm0.14$) & $0.51\pm0.21$ ($0.88\pm0.24$) & 0.679 \\
\end{tabular}
\end{ruledtabular}
\end{table}

\end{document}